\documentclass[aps,prd,amsmath,twocolumn,amssymbaps,showpacs]{revtex4-1}
\usepackage{graphicx}  
\usepackage{dcolumn}   
\usepackage{bm}        
\usepackage{amssymb}   
\usepackage{amsmath}   
\usepackage{draftcopy}
\usepackage{relsize} 
\usepackage{xfrac}    
\usepackage{slashed}

\usepackage{color}
\definecolor{dgreen}{cmyk}{1.,0.,1.,0.2}        
\definecolor{orange}{cmyk}{0.,0.353,1.,0.}    

\usepackage[bookmarks]{hyperref}

\newcommand{\di}{{\rm d}}

\newcommand{\be}{\begin{equation}}
\newcommand{\ee}{\end{equation}}                                                                               
\newcommand{\bea}{\begin{eqnarray}}
\newcommand{\eea}{\end{eqnarray}}

\begin{document}
\title{Charged rho superconductor in the presence of magnetic field and rotation}

\author{Gaoqing Cao}
\affiliation{School of Physics and Astronomy, Sun Yat-sen University, Guangzhou 510275.}
\date{\today}

\begin{abstract}
In this work, we mainly explore the possibility of charged rho superconductor (CRS) in the presence of parallel magnetic field and rotation within three-flavor Nambu--Jona-Lasino model. By following similar schemes as in the previous studies of charged pion superfluid (CPS), the CRS is found to be favored for both choices of Schwinger phase in Minkovski and curved spaces. Due to the stability of the internal spin structure, charged rho begins to condensate at a smaller threshold of angular velocity than charged pion for the given large magnetic fields. Even the axial vector meson condensation is checked -- the conclusion is that CRS is the robust ground state at strong magnetic field and fast rotation, which actually sustains to very large angular velocity.
\end{abstract}

\maketitle

\section{Introduction}

Nowadays, several extraordinary conditions can be realized in the terrestrial relativistic heavy ion collisions (HICs), such as strong electromagnetic (EM) field~\cite{Skokov:2009qp,Deng:2012pc,Guo:2019mgh,Xu:2020sui} and fast rotation~\cite{Liang:2004ph,Becattini:2016gvu,STAR:2017ckg}. Under such circumstances, the properties of quantum chromodynamics (QCD) system are quite interesting and attractive topics. Actually, magnetic field and rotation share some similar effect, thus the proposal of chiral magnetic effect is right followed by that of chiral vortical effect around 2008. These anomalous transport phenomena were intensively studied since then~\cite{Liao:2014ava,Kharzeev:2015znc,Huang:2015oca}, and recently a very important breakthrough has been acheived in the BES II experiment of STAR group~\cite{STAR:2020crk}. Nevertheless, along with the discoveries of magnetic catalysis effect at zero temperature~\cite{Gusynin:1994re,Gusynin:1994xp} and global polarization of $\Lambda$ hyperon in peripheral HICs~\cite{Karpenko:2016jyx,Li:2017slc,Niida:2018hfw}, some unexpected features emerge and still require proper explanations: the inverse magnetic catalysis effect~\cite{Bali:2011qj,Bali:2012zg} and the "sign puzzles" of the local polarizations~\cite{Niida:2018hfw,Becattini:2017gcx,Xia:2018tes,Becattini:2019ntv,Xia:2019fjf}. In some sense, the extreme conditions open a wide realm for the searching of new phases, such as CRS in pure magnetic field~\cite{Chernodub:2010qx,Chernodub:2011mc}, neutral pseudoscalar superfluid in parallel EM field~\cite{Cao:2015cka,Wang:2017pje,Wang:2018gmj,Cao:2020pjq} and CPS in parallel magnetic field and rotation (PMR)~\cite{Liu:2017spl}. Among others, the possibility of CRS was under fierce debate since its proposal~~\cite{Chernodub:2010qx,Chernodub:2011mc,Braguta:2011hq,Liu:2014uwa,Hidaka:2012mz,Bali:2017ian,Cao:2019res,Ding:2020jui}, mainly concerning the internal quark-antiquark effect on rho mesons.

Recently, the existence of CPS in PMR became also controversial according to the studies in Nambu--Jona-Lasinio (NJL) model, where one is immersed in the ambiguity of the definition for Schwinger phase~\cite{Cao:2019ctl,Chen:2019tcp}. The breaking effect of rotation on the internal spin structure of charged pion was checked in these works for choices of Schwinger phase in Minkovski (SPM) and curved (SPC) spaces, respectively. It turned out that CPS is never favored for SPC and only favored in the intermediate regime of angular velocity for SPM~\cite{Chen:2019tcp}. As mentioned in the conclusion of Ref.~\cite{Cao:2019ctl}, the spins of valence quark and antiquark are along the same direction in rho vector mesons; thus, the spin-up $\rho^+$ meson is stable in the presence of either strong magnetic field $B$ or large rotating angular velocity $\Omega$ along $z$ direction. Due to the mass reduction in $B$ and effective isospin chemical potential generated by $\Omega$, it is quite probable that CRS would occur in PMR and keep robust to very fast rotation. Similar to the electric field discussed in Ref.\cite{Cao:2015cka}, the rotation term breaks the semi-positivity of fermion determinante in the partition function. Therefore, the study in such setup is free from the constraint of Vafa-Witten (VW) theorem~\cite{Vafa:1984xg,Hidaka:2012mz}, which was previously adopted as the main point against CRS in pure magnetic field. To show the importance of rotation effect on CRS, we'd like to mention the interesting results found in Ref.\cite{Zhang:2018ome}: At finite isospin chemical potential $\mu_I$, CPS is always favored over CRS; but CRS  would finally manage to overwhelm CPS with $\Omega$ increasing.

After getting some intuitions from the Weinberg model in Sec.\ref{Weinberg model}, the paper keeps a similar structure as our previous work Ref.~\cite{Cao:2019ctl}. In Sec.\ref{model}, we present the formalism for $SU(3)$ NJL model in rotating frame with a parallel magnetic field, where the simplest forms of vector interactions are introduced to explore rho meson physics~\cite{Cao:2019res}. Then,  the quadratic coefficient in Ginzburg-Landau expansion will be evaluated analytically in Sec.\ref{coefficient} with the choice of SPM in Sec.\ref{ASPM} and  of SPC in Sec.\ref{ASPC}, respectively. Eventually, the numerical results will be illuminated in Sec.\ref{numerical}  to check the stability of QCD system against CRS and we give a simple conclusion in Sec.\ref{conclusions}. The natural units $c=\hbar=k_{\rm B}=1$ are used throughout.

\section{Intuitions from Weinberg model}\label{Weinberg model}
From the chiral effective Weinberg model~\cite{Weinberg:1968de} with pion and rho mesons the fundamental degrees of freedom, the Lagrangian density can be extended to the case with PMR ias
\begin{eqnarray}
{\cal L}&=&{1\over 2}\left({(D_\mu\boldsymbol{\pi})^\dagger\cdot D^\mu\boldsymbol{\pi}\over(1+{\boldsymbol{\pi}^\dagger\cdot\boldsymbol{\pi}\over f_\pi^2})^2}-{m_\pi^2\,\boldsymbol{\pi}^\dagger\cdot\boldsymbol{\pi}\over 1+{\boldsymbol{\pi}^\dagger\cdot\boldsymbol{\pi}\over f_\pi^2}}\right)-{1\over 4} \boldsymbol{\rho}_{\mu\nu}^\dagger\cdot\boldsymbol{\rho}^{\mu\nu}\nonumber\\
&&+{m_\rho^2\over2}\left[\boldsymbol{\rho}_\mu\!+\!{g_\rho\boldsymbol{\pi}\!\times\! D_\mu\boldsymbol{\pi}\over m_\rho^2(1\!+\!{\boldsymbol{\pi}^\dagger\cdot\boldsymbol{\pi}\over f_\pi^2})}\right]^\dagger\cdot\left[\boldsymbol{\rho}^\mu\!+\!{g_\rho\boldsymbol{\pi}\!\times\! D^\mu\boldsymbol{\pi}\over m_\rho^2(1\!+\!{\boldsymbol{\pi}^\dagger\cdot\boldsymbol{\pi}\over f_\pi^2})}\right]\nonumber\\
&&\pm i{e\over 2}F^{\mu\nu}{\rho}_\mu^\mp{\rho}_\nu^\pm-{B^2\over 2},
\end{eqnarray}
where chiral symmetry is nonlinearly realized through the term $1+{\boldsymbol{\pi}^\dagger\cdot\boldsymbol{\pi}\over f_\pi^2}$ in the denominators, and magnetic and rotation effects are encoded in the covariant derivative $D_\mu$. Neglecting all the self-interactions of pions for simplicity, the Lagrangian density is then reduced to
\begin{eqnarray}\label{WM}
{\cal L}&=&-{B^2\over 2}+{1\over 2}\left[{(D_\mu\boldsymbol{\pi})^\dagger\cdot D^\mu\boldsymbol{\pi}}-{m_\pi^2\,\boldsymbol{\pi}^\dagger\cdot\boldsymbol{\pi}}\right]-{1\over 4} \boldsymbol{\rho}_{\mu\nu}^\dagger\cdot\boldsymbol{\rho}^{\mu\nu}\nonumber\\
&&+{m_\rho^2\over2}\boldsymbol{\rho}_\mu^\dagger\cdot\boldsymbol{\rho}^\mu\pm i{e\over 2}F^{\mu\nu}{\rho}_\mu^\mp{\rho}_\nu^\pm+{g_\rho\over 2}\left[\boldsymbol{\rho}_\mu^\dagger\cdot({\boldsymbol{\pi}\times D^\mu\boldsymbol{\pi}})\right.\nonumber\\
&&\left.+(\boldsymbol{\pi}\times D_\mu\boldsymbol{\pi})^\dagger\cdot\boldsymbol{\rho}^\mu\right].
\end{eqnarray}
Here, the isovectors are defined in the electric charge eigenstates: $\boldsymbol{\pi}=(\pi^0,\pi^-,\pi^+)$ and $\boldsymbol{\rho}=(\rho^0,\rho^-,\rho^+)$, and we assume for convenience that  the rho vector mesons are in the spin eigenstates: $\rho_\mu=(\rho_t,\rho_\downarrow,\rho_0,\rho_\uparrow)$. Take $\rho$ mesons for example, the charge eigenstates are related to the isospin ones $\rho^{i}\ (i=1,2,3)$ as
$$\rho^0=\rho^3,\ \rho^\pm={\rho^1\mp i\rho^2\over{\sqrt{2}}},$$
and the spin eigenstates are defined by the Lorentz components $\rho_\mu\ (\mu=t,x,y,z)$ as
$$\rho_0=\rho_z,\ \rho_{\uparrow/\downarrow}={\rho_{x}\mp i\,\rho_{y}\over \sqrt 2}.$$ 

In the vacuum, the strength tensors of $\rho$ mesons are defined in a similar way as those of gauge fields in the $SU(2)$ Yang-Mills theory:
$${\rho}_{\mu\nu}^a\equiv \partial_\mu{\rho}_{\nu}^a-\partial_\nu{\rho}_{\mu}^a+g_\rho\epsilon^{abc}{\rho}_{\mu}^b{\rho}_{\nu}^c$$
with the coupling constant given by $g_\rho=\sqrt{2}m_\rho/f_\pi$~\cite{Weinberg:1968de}. These tensors can be rearranged in the charge eigenstates so that the magnetic effect can be introduced directly by changing $\partial_\mu$ to covariant derivative $D_\mu\equiv\partial_\mu+i\,qA_\mu$ with $q$ the particle charge. Then, we get the strength tensors of charge-definite $\boldsymbol{\rho}$ as
\begin{eqnarray}
{\rho}_{\mu\nu}^0&=&\partial_\mu{\rho}^0_\nu-\partial_\nu{\rho}^0_\mu+i\,g_\rho (\rho^-_\mu\rho^+_\nu-\rho^-_\nu\rho^+_\mu),\\
{\rho}_{\mu\nu}^\pm&=&D_\mu^\pm{\rho}^\pm_\nu-D_\nu^\pm{\rho}^\pm_\mu,\ D_\mu^\pm\equiv\partial_\mu\pm ieA_\mu\mp i\,g_\rho{\rho}^0_\mu,
\end{eqnarray}
where we find that ${\rho}_{\mu\nu}^\pm$ can be simply present in Abelian forms with the redefinition of the gauge field as $A_\mu-{g_\rho\over e}{\rho}^0_\mu$.
In accordance with the spin eigenstates, the corresponding covariant derivatives are related to the Lorentz components as:
$$D_0^{(\pm)}=D_z^{(\pm)},\ D_{\uparrow/\downarrow}^{(\pm)}=i{D_x^{(\pm)}\mp i\,D_y^{(\pm)}\over \sqrt 2}$$
for both neutral and charged $\rho$ mesons. Considering a constant magnetic field along $z$ direction, we choose the symmetric gauge for the vector potential: $A_\mu=(0,By/2,-Bx/2,0)$. Then, as introduced in Ref.~\cite{Lee:1962vm}, the strength tensor couplings to the EM field  in Eq.\eqref{WM} become explicitly
$$\pm i{e\over 2}F^{\mu\nu}{\rho}_\mu^\mp{\rho}_\nu^\pm=\pm {1\over 2}eB\left({\rho}_{\downarrow}^\mp{\rho}_\uparrow^\pm-{\rho}_{\uparrow}^\mp{\rho}_\downarrow^\pm\right).$$
Furthermore, according to the discussions in Ref.\cite{Chen:2017xrj,Jiang:2016wvv}, the effect of rotation along $z$ direction can be simply introduced through the modification of temporal derivative $\partial_t$ to 
$$D_t=\partial_t-i\,\Omega\left(\hat{L}_z+\hat{S}_z\right),$$
where $\hat{L}_z\equiv-i(x\partial_y-y\partial_x)$ and $\hat{S}_z$ are the orbital angular momentum (OAM) and spin operators, respectively.

Now, without applying any boundary condition to a cylindrical system with radius $R$, the diagonal kinetic parts of the Lagrangian can be expressed explicitly on the basis of energy $k_4$, momentum $k_3$, Landau level $n$ and OAM quantum number $l$ as
\begin{widetext}
\begin{eqnarray}
{\cal L}_k&=&-{1\over 2}{{\pi}^0_{-l}\left[(k_4-i\,\Omega l)^2+{k}_l^2+{k}_3^2+m_\pi^2\right]{\pi}^0_l}-{1\over 2}{{\pi}^\pm_{n,\pm l}\left[(k_4\pm i\,\Omega l)^2+(2n+1)|eB|+k_3^2+m_\pi^2\right]{\pi}^\mp_{n,\mp l}}\nonumber\\
&&-{1\over 2} {\rho}^{0}_{-s,-l}\left[(k_4-i\,\Omega (l+s))^2+{{k}_l^2}+{k}_3^2+m_\rho^2\right]{\rho}^{0}_{s,l}+{1\over 2}{\rho}^{0}_{t,-l}\left[(k_4-i\,\Omega (l+s))^2+{k}_l^2+{k}_3^2+m_\pi^2\right]{\rho}^{0}_{t,l}\nonumber\\
&&-{1\over 2} {\rho}^{\pm}_{-s,n,\pm l}\left[(k_4-i\,\Omega (\mp l+s))^2+(2n+1\pm 2s)|eB|+{k}_3^2+m_\rho^2\right]{\rho}^{\mp}_{s,n,\mp l}\nonumber\\
&&+{1\over 2} {\rho}^{\pm}_{t,n,\pm l}\left[(k_4\pm i\,\Omega l)^2+(2n+1)|eB|+{k}_3^2+m_\rho^2\right]{\rho}^{\mp}_{t,n,\mp l},\label{Lk}
\end{eqnarray}
where particularly the summations over $n,l,s$ should be understood with $s=-1,0,1$, $n\geq0$, $l\in(-\infty,\infty)$ for neutral particles and $l\in(-n,[\mathcal{N}]-n)$ for charged ones~\cite{Chen:2017xrj}. Note that the OAM is given by $-l$ for negative charged particle and $[\mathcal{N}]\equiv\left[{|qB| R^2\over 2}\right]$ is the number of magnetic flux quantization. Especially, we choose the simplest Lorentz gauge $D_\mu\boldsymbol{\rho}^\mu=0$ for $\rho$ mesons, then the commutations $[D_\mu,D_\nu]$ from $\boldsymbol{\rho}_{\mu\nu}^\dagger\cdot\boldsymbol{\rho}^{\mu\nu}$ give rise to extra kinetic terms the same as the strength tensor couplings. The left particle coupling parts of the Lagrangian involve the self-interactions of $\rho$ mesons, which are quite the same as those of $W/Z$ bosons in the electroweak theory, and the $\rho\pi\pi$ interactions whose explicit forms can be illuminated as
\begin{eqnarray}
{\cal L}_{\rho\pi\pi}&=&g_\rho\left\{\left[\left({\rho}^{-}_{\downarrow\uparrow,n,-l}{\pi}^+_{n', l'}-{\rho}^{+}_{\downarrow\uparrow,n,l}{\pi}^-_{n',-l'}\right)\left(\pm {k_{l''}''}{\pi}^0_{l''\mp 1}\right)/{\sqrt 2}-\left({\rho}^{-}_{0,n,-l}{\pi}^+_{n', l'}-{\rho}^{+}_{0,n,l}{\pi}^-_{n',-l'}\right)k_3''{\pi}^0_{l''}-i\left({\rho}^{-}_{t,n,-l}{\pi}^+_{n', l'}\right.\right.\right.\nonumber\\
&&\left.\left.-{\rho}^{+}_{t,n,l}{\pi}^-_{n',-l'}\right)(k_4-i\,\Omega l''){\pi}^0_{l''}\right]-\left[\left({\rho}^{-}_{\downarrow,n,-l}\sqrt{(n''+1)|eB|}{\pi}^+_{n''+1, l''-1}+{\rho}^{+}_{\downarrow,n,l}\sqrt{n''|eB|}{\pi}^-_{n''- 1,-l''-1}\right){\pi}^0_{l''- 1}\right.\nonumber\\
&&-\left({\rho}^{-}_{\uparrow,n,-l}\sqrt{n''|eB|}{\pi}^+_{n''-1, l''+1}+{\rho}^{+}_{\uparrow,n,l}\sqrt{(n''+1)|eB|}{\pi}^-_{n''+ 1,-l''+1}\right){\pi}^0_{l''+ 1}-\left({\rho}^{-}_{0,n,-l}k_3'{\pi}^+_{n', l'}-{\rho}^{+}_{0,n,l}k_3'{\pi}^-_{n',-l'}\right){\pi}^0_{l''}\nonumber\\
&&\left.-i\left({\rho}^{-}_{t,n,-l}(k_4-i\,\Omega l'){\pi}^+_{n', l'}-{\rho}^{+}_{t,n,l}(k_4+i\,\Omega l'){\pi}^-_{n',-l'}\right){\pi}^0_{l''}\right]+\left[{\rho}^{0}_{\downarrow,l}\left({\pi}^-_{n',-l'}\sqrt{(n''+1)|eB|}{\pi}^+_{n''+1, l''-1}\right.\right.\nonumber\\
&&\left.+{\pi}^+_{n', l'}\sqrt{n''|eB|}{\pi}^-_{n''- 1,-l''-1}\right)-{\rho}^{0}_{\uparrow,l}\left({\pi}^-_{n',-l'}\sqrt{n''|eB|}{\pi}^+_{n''-1, l''+1}+{\pi}^+_{n', l'}\sqrt{(n''+1)|eB|}{\pi}^-_{n''+ 1,-l''+1}\right)\nonumber\\
&&\left.\left.-{\rho}^{0}_{0,l}\left({\pi}^-_{n',-l'}k_3''{\pi}^+_{n'', l''}-{\pi}^+_{n', l'}k_3''{\pi}^-_{n'',-l''}\right)-i{\rho}^{0}_{t,l}\left({\pi}^-_{n',-l'}(k_4-i\,\Omega l''){\pi}^+_{n'', l''}-{\pi}^+_{n', l'}(k_4+i\,\Omega l''){\pi}^-_{n'',-l''}\right)\right]\right\}.\label{LRPP}
\end{eqnarray}
\end{widetext}
Here, as before, the summations over the quantum numbers of all the relevant particles should be understood. We note that the coordinate integrations haven't been carried out yet in Eq.\eqref{LRPP}, that is why there seem no connections among the quantum numbers of the interacting particles.

In the following, we skip the complicated interaction parts and only focus on the kinetic parts of the Lagrangian to get some physical intuitions. From Eq.\eqref{Lk}, we surprisingly notice that rotation even induces effective chemical potentials for neutral pion and rho, thus the $\pi^0\ (\rho^0)$ condensation is expected when $|\Omega l|>m_{\pi}\ (|\Omega(l+s)|>m_{\rho})$. Moreover, in the presence of a magnetic field, the $\pi^0$ seems much easier to condense than $\pi^\pm$ with the condition for the latter: $|\Omega l|>\sqrt{m_\pi^2+|eB|}$. All these puzzles can be consistently solved when we combine the restriction of causality together with boundary conditions~\cite{Chen:2017xrj,Davies:1996ks,Ambrus:2015lfr}. As a matter of fact, causality constrains the angular velocity to $\Omega\leq1/R$ and the Dirichlet boundary conditions, requiring the wave functions to vanish at the boundary $R$, discretizes the transverse momentum such that $|k_{l}|>(|l|+2)/R$ for each $l$. Thus, the excitation energy $E_0=({k}_l^2+{k}_3^2+m^2)^{1/2}$ of the neutral particle satisfies
$$E_0>|k_{l}|>|\Omega(l+s)|,$$
which eventually prevents any accumulation of $\pi^0$ or $\rho^0$ meson.

Next, we discuss a bit more about the effect of boundary condition on $\pi^\pm$ in a background magnetic field. For $\pi^+$ with $l>0$, the excitation energy for transverse dynamics is~\cite{Liu:2017spl} 
$$E_{nl}(eB)\equiv \sqrt{m_\pi^2+|eB|(2\lambda_l^n+1)}$$
with the boundary condition
	\begin{eqnarray}
	_1F_1(-\lambda_l^n,l+1,\mathcal{N})=0\nonumber
	\end{eqnarray} 
and the quasi Landau levels $0\leq\lambda_l^0<\lambda_l^1<\lambda_l^2<\dots$. We've checked numerically that there is always a window of $l$ satisfying $l^2>2(2\lambda_l^0+1)\mathcal{N}$ when $\mathcal{N}\gtrsim 7$, which means the unstable condition $\Omega l>E_{0l}(eB)$ can be realized for large enough $\Omega$ and $eB$. This is consistent with CPS found in Ref.~\cite{Liu:2017spl}. However, one problem is still left: if we don't artificially set $l\leq [\mathcal{N}]$ as in Ref.~\cite{Liu:2017spl}, it seems that the lowest total particle energy $E_{0l}^\Omega=E_{0l}(eB)-\Omega l$ is not bound from below, which would cause a disaster of infinite condensate density according to Ref.~\cite{Liu:2017spl}. Actually, the answer is that $\lambda_l^0$ becomes quite large for large $l$, which then makes sure that $E_{0l}^\Omega>-\infty$ in the limit $l\rightarrow\infty$. 

\begin{figure}[!htb]
	\centering
	\includegraphics[width=0.42\textwidth]{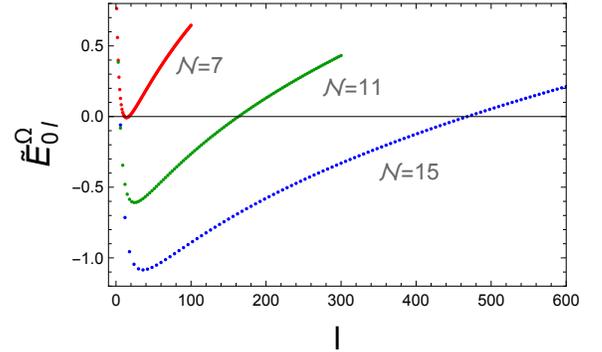}
	\caption{The dimensionless energy $\tilde{E}_{0l}^\Omega$ as a function of the orbital angular momentum $l$ for different values of $\mathcal{N}$.}\label{DE}
\end{figure}

\begin{figure}[!htb]
	\centering
	\includegraphics[width=0.42\textwidth]{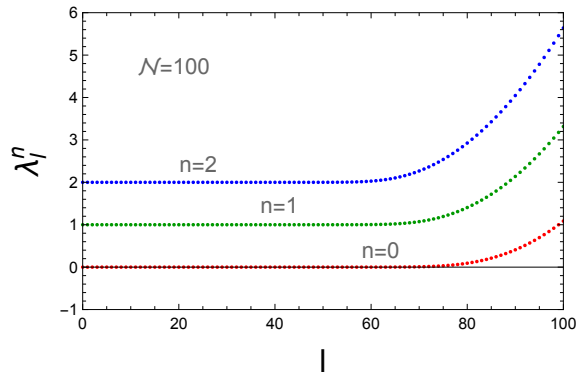}
	\caption{The three lowest quasi Landau levels $\lambda_l^n\ (n=0,1,2)$ as functions of the orbital angular momentum $l$ for $\mathcal{N}=100$.}\label{lambda_l}
\end{figure}

For the purpose of intuition, we show the scaled dimensionless energy 
\begin{eqnarray}
\tilde{E}_{0l}^\Omega\equiv{\sqrt{|eB|(2\lambda_l^0+1)}-{l\over R}\over \sqrt{|eB|}}=\sqrt{2\lambda_l^0+1}-{l\over\sqrt{2\mathcal{N}}}\nonumber
\end{eqnarray}
for different values of $\mathcal{N}$ in Fig.\ref{DE}, where positive features can always be identified at large $l$. And with the increasing of $\mathcal{N}$, that is, the enhancement of $eB$ for a given $R$, more $\pi^+$ can be condensed as more are trapped in the system. It should be pointed out that finite lower boundaries exist for the total energies of any charged particles.
In pure magnetic field, it was found that the degeneracy of $l$ is automatically restricted to $\leq[\mathcal{N}]$ for the lowest Landau level when the boundary condition is applied~\cite{Chen:2017xrj}. We even check in advance that the degeneracy decreases with the Landau level $n$, see the plains in Fig.\ref{lambda_l}. Nevertheless, for the case $\mathcal{N}=15$ in Fig.\ref{DE}, the much wider unstable window of $l$ disfavors the use of the artificial upper bound $[\mathcal{N}]$ for $l$ when large angular velocity ($\Omega\lesssim1/R$) is involved.

Finally, turn to the charged $\rho$ vector mesons, the story is quite different because of their non-vanishing spins. In the presence of PMR, the effective mass of $\rho^+_{\uparrow}$ or $\rho^-_\downarrow$ decreases as $\sqrt{m_\rho^2+|eB|(2\lambda_l^0-1)}$ on one hand~\cite{Chernodub:2010qx}, the effective isospin chemical potential increases as $\Omega(l+1)$ on the other hand. Then, it seems that $\rho^+_{\uparrow}$ condensation would overwhelm the $\pi^+$ condensation to be the true ground state when $\Omega$ is large enough that 
$$\Omega(l+1)-\sqrt{m_\rho^2+|eB|(2\lambda_l^0-1)}>\Omega l-E_{0l}(eB)>0.$$
 As the VW theorem might forbid the decreasing of composite $\rho^+_{\uparrow}$ mass to zero in pure magnetic field~\cite{Hidaka:2012mz,Bali:2017ian,Cao:2019res,Ding:2020jui}, the estimation of the $\rho^+_{\uparrow}$ mass is not correct at all for large $B$ in the point particle picture. However, as mentioned in the introduction, $\Omega$ invalids the proof of the theorem thus the isospin chemical potential effect of $\Omega(l+1)$ can still be qualitatively correct in the point particle picture. As a strong support of this point, we'd like to mention that the isospin effect was first discussed in chiral perturbation theory with pions the fundamental degrees of freedom~\cite{Son:2000xc}, and the proposed CPS was well verified by the effective NJL model~\cite{He:2005nk,Cao:2016ats} and lattice QCD simulations~\cite{Kogut:2002zg,Brandt:2017oyy} with quarks the fundamental degrees of freedom.

\section{Nambu--Jona-Lasinio model in rotating frame}\label{model}

In order to explore the possibility of charged rho condensation more realistically, we adopt the $SU(3)$ NJL model with $u, d$ and $s$ quarks the fundamental degrees of freedom~\cite{Klevansky:1992qe}. In the rotating frame, the action of the system can be conveniently given in curved spacetime by 
\begin{eqnarray}
{\cal S}=\int d^4x \sqrt{-\det(g_{\mu\nu})}{\cal L}(\bar{\psi},\psi),
\end{eqnarray}
where the Lagrangian density can be extended from the usual one~\cite{Klevansky:1992qe,Klimt:1989pm} to
\begin{eqnarray}
{\cal L}_{\rm NJL}&=&\bar\psi(i\slashed{D}-m_0)\psi+G_S\sum_{a=0}^8[(\bar\psi\lambda^a\psi)^2+(\bar\psi i\gamma_5\lambda^a\psi)^2]\nonumber\\
&&+{\cal L}_6-G_V\left[\left(\bar\psi\gamma^\mu\tau^a\psi\right)^2+\left(\bar\psi \gamma^\mu\gamma_5\tau^a\psi\right)^2\right]\nonumber\\
&{\cal L}_6&=-K\sum_{s=\pm}{\rm Det}\bar\psi\Gamma^s\psi
\end{eqnarray} 
by further adopting the four fermion vector interaction channels with coupling constant $G_V$. Compared to the two-flavor NJL model, the advantage of three-flavor NJL model is that there the vacuum superconductivity or CRS cannot happen in pure magnetic field~\cite{Cao:2019res} which is consistent with lattice QCD simulations~\cite{Hidaka:2012mz,Bali:2017ian,Ding:2020jui}.

In the Lagrangian, $\psi=(u,d,s)^T$ represents the three-flavor quark field and $m_0={\rm diag}(m_{\rm 0u},m_{\rm 0d},m_{\rm 0s})$ is the current quark mass matrix. The longitudinal and transverse covariant derivatives with PMR effect are defined, for symmetric gauge, respectively as 
$$D_0=\partial_t-i\Omega \left(\hat{L}_z+\hat{S}_z\right),~D_3=\partial_z$$
and
$$D_1=\partial_x+iQ\,By/2,~D_2=\partial_y-iQ\,Bx/2$$
with the charge matrix $Q={\rm diag}(q_{\rm u},q_{\rm d},q_{\rm s})$. For the four-fermion interaction terms, $\lambda^0=\sqrt{2\over3}I$ and Gell-Mann matrices $\lambda^i~(i=1,\dots,8)$ are defined in three-flavor space, so the extra diagonal terms $(\bar\psi\lambda^{3}\psi)^2$ and $(\bar\psi\lambda^{8}\psi)^2$ allow mass splitting among all the flavors in contrary to the two-flavor case~\cite{Cao:2019res}. The $U_A(1)$ symmetry violating term ${\cal L}_6$~\cite{tHooft:1976snw} only involves scalar-pseudoscalar channels with the determinant defined in flavor space, $\Gamma^\pm=1\pm\gamma_5$ and $K$ the coupling constant. Now, we only consider nonzero chiral condensations $\sigma_{\rm i}\equiv\langle\bar{\psi}^i{\psi}^i\rangle$, where the correspondence between the Arabian denotations $i=1,2,3$ and the more explicit Latin ones ${\rm f=u,d,s}$ should be understood for the flavors. The six fermion interactions in ${\cal L}_6$ can be reduced to effective four fermion ones in Hartree approximation~\cite{Klevansky:1992qe}, then the Lagrangian density only involves four fermion effective interactions:
\begin{widetext}
	\begin{eqnarray}
	\!\!\!{\cal L}_{\rm NJL}^4=\bar\psi(i\slashed{D}-m_0)\psi+\!\!\sum_{a,b=0}^8\!\left[G_{ab}^-(\bar\psi\lambda^a\psi)(\bar\psi\lambda^b\psi)\!+\!G_{ab}^+(\bar\psi i\gamma_5\lambda^a\psi)(\bar\psi i\gamma_5\lambda^b\psi)\right]\!-\!G_V\!\!\left[\left(\bar\psi\gamma^\mu\tau^a\psi\right)^2\!+\!\left(\bar\psi \gamma^\mu\gamma_5\tau^a\psi\right)^2\right],
	\end{eqnarray}
	where the non-vanishing elements of the symmetric coupling matrices $G^\pm$ are given by~\cite{Klevansky:1992qe}
	\begin{eqnarray}
	&&G_{00}^\mp=G_S\mp {K\over3}\sum_{\rm f=u,d,s}\sigma_{\rm f},~G_{11}^\mp=G_{22}^\mp=G_{33}^\mp=G_S\pm {K\over2}\sigma_s,~G_{44}^\mp=G_{55}^\mp=G_S\pm {K\over2}\sigma_{\rm d},~G_{66}^\mp=G_{77}^\mp=G_S\pm {K\over2}\sigma_{\rm u},\nonumber\\
	&&G_{88}^\mp=G_S\mp {K\over6}(\sigma_s-2\sigma_{\rm u}-2\sigma_{\rm d}),~G_{08}^\mp=\mp {\sqrt{2}K\over12}(2\sigma_s\!-\!\sigma_{\rm u}\!-\!\sigma_{\rm d}),~G_{38}^\mp=-\sqrt{2}G_{03}^\mp=\mp {\sqrt{3}K\over6}(\sigma_{\rm u}\!-\!\sigma_{\rm d}).
	\end{eqnarray}

In the case $\langle\sigma_i\rangle\neq0$, the inverse quark propagators of different flavors are given by the introductions of the dynamical masses and covariant derivatives as
\bea
S_{i}^{-1}(x,x')=i\slashed{D}-m_i,\ m_i=m_{0i}-4G_S\sigma_i+K\sum_{jk}\!\epsilon_{ijk}^2\sigma_j\sigma_k.
\eea
By using the eigenfunction reconstruction method, we have given the propagator of a  fermion with positive or negative charge~\cite{Cao:2019ctl}; so the $u$ and $d/s$ quark propagators are respectively
\begin{eqnarray}\label{propogator}
\!\!\!S_{\rm u}(x,x')&=&\sum_{n=0}^\infty\sum_l\int\int{dp_0dp_z\over(2\pi)^2}{i~e^{-ip_0(t-t^\prime)+ip_z(z-z^\prime)}\over\left({p}_0^{l+}\right)^2-(\varepsilon_n^u)^2+i\epsilon}
	\Bigg\{\left[{\cal P}_\uparrow\chi_{n,l}^+(\theta,r)\chi_{n,l}^{+*}(\theta',r')+{\cal P}_\downarrow\chi_{n-1,l+1}^+(\theta,r)\chi_{n-1,l+1}^{+*}(\theta',r')\right]\nonumber\\
	&&\!\!\!\!\!\left(\gamma^0{p}_0^{l+}-\gamma^3p_z+m_u\right)-\left[{\cal P}_\uparrow\chi_{n,l}^+(\theta,r)\chi_{n-1,l+1}^{+*}(\theta',r')+{\cal P}_\downarrow\chi_{n-1,l+1}^+(\theta,r)\chi_{n,l}^{+*}(\theta',r')\right]\sqrt{2n |qB|}\gamma^2\Bigg\}_{q\rightarrow q_{\rm u}},\label{Su}\\
\!\!\!S_{\rm d/s}(x,x')&=&\sum_{n=0}^\infty\sum_l\int_{-\infty}^{\infty}{dp_0dp_z\over(2\pi)^2}{i~e^{-ip_0(t-t^\prime)+ip_z(z-z^\prime)}\over\left({p}_0^{l-}\right)^2-(\varepsilon_n^{d/s})^2+i\epsilon}
	\Bigg\{\left[{\cal P}_\uparrow\chi_{n-1,l-1}^{-}(\theta,r)\chi_{n-1,l-1}^{-*}(\theta',r')+{\cal P}_\downarrow\chi_{n,l}^{-}(\theta,r)\chi_{n,l}^{-*}(\theta',r')\right]\nonumber\\
	&&\!\!\!\!\!\!\!\left(\gamma^0{p}_0^{l-}-\gamma^3p_z+m_{\rm d/s}\right)+\left[{\cal P}_\uparrow\chi_{n-1,l-1}^{-}(\theta,r)\chi_{n,l}^{-*}(\theta',r')+{\cal P}_\downarrow\chi_{n,l}^{-}(\theta,r)\chi_{n-1,l-1}^{-*}(\theta',r')\right]\sqrt{2n |qB|}\gamma^2\Bigg\}_{q\rightarrow q_{\rm d}},\label{Sds}
\end{eqnarray}
\end{widetext}	
where ${p}_0^{ls}=p_0+\Omega \left(l+s{1\over2}\right)$, the dispersion relations $\varepsilon_n^i=(p_z^2+2n|q_iB|+m_i^2)^{1/2}$ and ${\cal P}_{\uparrow/\downarrow}={1\over2}(1\pm\sigma^{12})$ are the spin projectors. Here, the normalized auxiliary functions are defined for positive and negative charged particles respectively as
\begin{eqnarray}
\!\!\!\!\!\!\!\chi_{n,l}^+(\theta,r)&=&\left[{|qB|\over2\pi}{ n!\over(n\!+\!l)!}\right]^{1\over2}{e^{i\, l\theta}}~\tilde{r}^le^{-\tilde{r}^2/2}L_n^l\left(\tilde{r}^2\right),\\
\!\!\!\!\!\!\!\chi_{n,l}^{-}(\theta,r)&=&\left[{|qB|\over2\pi}{ n!\over(n\!-\!l)!}\right]^{1\over2}{e^{i\, l\theta}}~\tilde{r}^{-l}e^{-\tilde{r}^2/2}L_n^{-l}\left(\tilde{r}^2\right),
\end{eqnarray}
where the dimensionless radius $\tilde{r}^2=|qB|r^2/2$ and the Laguerre polynomial $L_n^l(x)$ is nonvanishing only for $n\ge0$. Actually, in a rotating system, a boundary must be applied due to causality, then the propagators would no longer keeps the forms of Eqs.\eqref{Su} and \eqref{Sds}. But for convenience, we still adopt these forms and constrain the OAM as $l\in\left[-n,\mathcal{N}-n\right]$~\cite{Chen:2019tcp}. Armed with that, the quark masses can be evaluated through the gap equations given by the self-consistent definitions of chiral condensations as:
\begin{eqnarray}
\sigma_i\equiv\langle\bar{\psi}^i{\psi}^i\rangle=-{i\over V_4}{\rm Tr}~S_{i}.
\end{eqnarray}
By adopting vacuum regularization, the explicit form of the gap equations are
\begin{widetext}
\begin{eqnarray}
-\sigma_{\rm f}
&=&N_c{{m_{\rm f}}^3\over2\pi^2}\Big[\tilde{\Lambda}_{\rm f}\Big({1+\tilde{\Lambda}_{\rm f}^2}\Big)^{1\over2}-\ln\Big({\tilde\Lambda_{\rm f}}
+\Big({1+\tilde{\Lambda}_{\rm f}^2}\Big)^{1\over2}\Big)\Big]+N_c{m_{\rm f}\over4\pi^2}\int_0^\infty {ds\over s^2}e^{-{m_{\rm f}}^2s}\left({q_{\rm f}Bs
	\over\tanh(q_{\rm f}Bs)}-1\right)\nonumber\\
&&-N_cm_{\rm f}\sum_{n=0}^{n_{\rm max}}{1\over S}\sum_{l=0}^{\mathcal{N}^{{\rm f}}}\int_{-\infty}^{\infty}{dp_z\over\pi}{\alpha_n\over\varepsilon_n^{{\rm f}}}
\Big[f(\varepsilon_n^{{\rm f}}+\Omega_{nl})+f(\varepsilon_n^{{\rm f}}-\Omega_{nl})\Big]\label{mgap}
\end{eqnarray}
\end{widetext}
with the reduced cutoff $\tilde{\Lambda}_{\rm f}={\Lambda/m_{\rm f}}$. Compared to that given in Ref.~\cite{Cao:2019ctl} but implicitly implied, the Landau levels are cut off by $n_{\rm max}~(\ll N_{\rm f})$ here, which was checked to be a good approximation for large $B$ and $\Omega=0$.

For a continuous transition, the effective potential can be expressed as the form in Ginzburg-Landau (GL) theory
\begin{eqnarray}\label{thermo}
V_{\rm eff}(\sigma_{\rm f},\Delta)=V_{\rm eff}(\sigma_{\rm f},0)+{\cal A}~\Delta^2+{\cal B}~\Delta^4+\dots,
\end{eqnarray}
where $V_{\rm eff}(\sigma_{\rm f},0)$ is the corresponding thermodynamic potential giving rise to the gap equations in Eq.\eqref{mgap}. We note that $\Delta$ can be an order parameter for any kind of mesonic superfluid or superconductor with the relevant coefficients ${\cal A}$ and ${\cal B}$ determined by their interactions with quarks. To avoid too much complexity, we assume the transition is solely determined by the quadratic one ${\cal A}$~\cite{Cao:2019ctl}: If ${\cal A}<0$, the meson condensation is favored; and if ${\cal A}_a<{\cal A}_b<0$, we would assume meson $a$ is more preferred to condensate than meson $b$. We've discovered previously that ${\cal A}$ is almost consistent with the inverse mesonic propagator in random phase approximation (RPA) except for some subtle discussions on the Schwinger phases of charged mesons~\cite{Cao:2019ctl,Chen:2019tcp,Cao:2015xja}. As illuminated in Ref.~\cite{Cao:2019res}, the bare form of the coefficient is given by
$${\cal A}={1\over4G}+\Pi$$
with the polarization function defined through the fermion loop as
\begin{equation}\label{PF}
\Pi={i\over V_4}{\rm Tr}\left[S(x,y)\Gamma_{M*}S(y,x)\Gamma_{M}e^{-i\Phi_M}\right].
\end{equation}
Here, $V_4$ is the space-time volume, the trace should be taken over the internal and coordinate spaces, and $e^{-i\Phi_M}$ is the compensated Schwinger phase.

\begin{widetext}
	\section{Calculations of the quadratic coefficient}\label{coefficient}

This section is mainly devoted to calculating the quadratic efficient explicitly. The interaction vertices between quarks and (pseudo-)scalar and vector mesons have been listed in Ref.~\cite{Cao:2019res} as:
	\bea
	\Gamma_{\sigma/\sigma^*}=-1,~\Gamma_{\pi^0/{\pi^0}^*}=-i\gamma^5\tau_3,~\Gamma_{\pi_\pm}=-i\gamma^5\tau_\pm,~\Gamma_{\bar{\omega}_\mu/\bar{\omega}_\mu^*}=\bar{\gamma}_\mu^\pm,~\Gamma_{\bar{\rho}_{0\mu}/\bar{\rho}_{0\mu}^*}=\bar{\gamma}_\mu^\pm\tau_3,~\Gamma_{\bar{\rho}_{\pm\mu}}=\bar{\gamma}_\mu^\pm\tau_\pm,
	\eea	
	where $\bar{\gamma}^{\pm}_\mu=(\gamma_0,{\gamma_1\pm i\gamma_2\over \sqrt{2}},{\gamma_1\mp i\gamma_2\over \sqrt{2}},\gamma_3)$. In the following, we mainly focus on the $\bar{\rho}^+_1$ mode, that is, the rho meson with spin long the magnetic field. The insertion of the fermion propagators from Eq.\eqref{mgap} into the polarization function Eq.\eqref{PF}  is explicitly
	
	\begin{eqnarray}
	\Pi_{\bar{\rho}^+_1}&=&{-i\over S}\sum_{n=0}^{n_{\rm max}}\sum_l\sum_{n'=0}^{n_{\rm max}}\sum_{l'}\int_{-\infty}^{\infty}{dp_0\over2\pi}\int_{-\infty}^{\infty}{dp_z\over2\pi}
	~{\rm Tr}\Bigg\{\left[{\cal P}_\uparrow\chi_{n,l}^+(\theta,r)\chi_{n,l}^{+*}(\theta',r')+{\cal P}_\downarrow\chi_{n-1,l+1}^{+}(\theta,r)\chi_{n-1,l+1}^{+*}(\theta',r')\right]\nonumber\\
	&&\ \ \ \ \times\left(\gamma^0{p}_0^{l+}-\gamma^3p_z+m_u\right)-\left[{\cal P}_\uparrow\chi_{n,l}^{+}(\theta,r)\chi_{n-1,l+1}^{+*}(\theta',r')+{\cal P}_\downarrow\chi_{n-1,l+1}^{+}(\theta,r)\chi_{n,l}^{+*}(\theta',r')\right]
	\gamma^2\sqrt{2n q_uB}\Bigg\}\nonumber\\
	&&\times\Bigg\{-\left[{\cal P}_\downarrow\chi_{n'-1,l'-1}^{-}(\theta',r')\chi_{n'-1,l'-1}^{-*}(\theta,r)+{\cal P}_\uparrow\chi_{n',l'}^{-}(\theta',r')\chi_{n',l'}^{-*}(\theta,r)\right]\left(\gamma^0{p}_0^{l'-}-\gamma^3p_z-m_d\right){\gamma^1+i\gamma^2\over \sqrt{2}}\nonumber\\
	&&\ \ \ \ \  -\left[{\cal P}_\downarrow\chi_{n'-1,l'-1}^{-}(\theta',r')\chi_{n',l'}^{-*}(\theta,r)+{\cal P}_\uparrow\chi_{n',l'}^{-}(\theta',r')\chi_{n'-1,l'-1}^{-*}(\theta,r)\right]\gamma^2\sqrt{2n'|q_dB|}{\gamma^1-i\gamma^2\over \sqrt{2}}\Bigg\}{\gamma^1-i\gamma^2\over \sqrt{2}}\nonumber\\
	&&\times{e^{-i\Phi}\over\left[\left({p}_0^{l+}\right)^2-(\varepsilon_n^u)^2\right]\left[\left({p}_0^{l'-}\right)^2-(\varepsilon_{n'}^d)^2\right]},
	\end{eqnarray}
	where the trace is over the Dirax and the coordinate spaces. By completing the trace over the Dirac space, the expression becomes quite simple:
	\begin{eqnarray}\label{A1}
	\Pi_{\bar{\rho}^+_1}&=&{-4N_ci\over S}\sum_{n=0}^{n_{\rm max}}\sum_l\sum_{n'=0}^{n_{\rm max}}\sum_{l'}\sum_{r,r'}\sum_{\theta,\theta'}\int_{-\infty}^{\infty}{dp_0\over2\pi}\int_{-\infty}^{\infty}{dp_z\over2\pi}
	{e^{-i\Phi}\over\left[\left({p}_0^{l+}\right)^2-(\varepsilon_n^u)^2\right]\left[\left({p}_0^{l'-}\right)^2-(\varepsilon_{n'}^d)^2\right]}\left({p}_0^{l+}{p}_0^{l'-}-p_z^2-m_um_d\right)\nonumber\\
	&&\chi_{n,l}^{+}(\theta,r)\chi_{n,l}^{+*}(\theta',r')\chi_{n',l'}^{-}(\theta',r')\chi_{n',l'}^{-*}(\theta,r),
	\end{eqnarray}
	where we define $\sum_{r,r'}=\int_0^\infty rdr\int_0^\infty r'dr'$ and $\sum_{\theta,\theta'}=\int_0^{2\pi}d\theta\int_0^{2\pi}d\theta'$. Consistent with the form in pure magnetic field~\cite{Cao:2019res}  but different from that of charged pion~\cite{Cao:2019ctl}, only the term with numerator independent of Landau levels survives and there is only one kind of combination of $\chi^{+}\chi^{+*}$ and $\chi^{-}\chi^{-*}$. By the way, for the corresponding axial vector $\bar{a}^+_1$ with interaction indices $\Gamma_{\bar{a}^+_1}=i\gamma^5\Gamma_{\bar{\rho}_1^+}$, the polarization function is the same as that of $\bar{\rho}^+_1$ except that the sign of the mass term is changed, that is, $-m_um_d\rightarrow+m_um_d$. We've shown in Ref.~\cite{Cao:2019res} that the contribution of the mass term is negative to $\Pi_{\bar{\rho}^+_1}$, hence $m_{\bar{a}^+_1}>m_{\bar{\rho}^+_1}$ in the chiral symmetry breaking phase. Then it turns out that the $\bar{a}^+_1$ superconductor is neither favored in magnetic field,  but it still needs to be checked in PMR where this term can be positive for $\bar{\rho}^+_1$.
	
\subsection{For Schwinger phase in Minkovski space}\label{ASPM}
	To calculate Eq.\eqref{A1} further, we choose the Schwinger phase of the form in Minkovski space, that is, $\Phi_M=e\int_x^y{A}_\mu(z) d{z}_\mu={eB\over2}\sin(\theta-\theta')rr'$. Then, the integrals over the polar angles can be completed to give
	\begin{eqnarray}\label{A2}
	\Pi_{\bar{\rho}^+_1}&=&{-16N_ci\over S}\sum_{n,n'=0}^{n_{\rm max}}\sum_{l=0}^{\mathcal{N}_u}\sum_{l'=0}^{\mathcal{N}_d}\sum_{r,r'}\int_{-\infty}^{\infty}{dp_0\over2\pi}\int_{-\infty}^{\infty}{dp_z\over2\pi}
	{e^{-{eB\over 4}(r^2+{r'}^2)}\left({p}_0^{(l-n)+}{p}_0^{(n'-l')-}-p_z^2-m_um_d\right)\over\left[\left({p}_0^{(l-n)+}\right)^2-(\varepsilon_n^u)^2\right]\left[\left({p}_0^{(n'-l')-}\right)^2-(\varepsilon_{n'}^d)^2\right]}{n!n'!\over l!l'!}\left({q_uB\over 2}\right)^{l-n+1}\nonumber\\
	&&\left({|q_dB|\over 2}\right)^{l'-n'+1}J_{l+l'-n-n'}\left({eB\over2}rr'\right)(rr')^{l+l'-n-n'}{F}_{nl,n'l'}(q_uB,|q_dB|;r,r'),
	\end{eqnarray}
	where the auxiliary function $F$ is defined as
	\begin{eqnarray}
	F_{nl,n'l'}(q_uB,|q_dB|;r,r')\equiv \prod_{x=r,r'}L_n^{l-n}\left({q_uB ~x^2\over2}\right)L_{n'}^{l'-n'}\left({|q_dB| ~x^2\over2}\right).
	\end{eqnarray}	
	Here, we find that the LLL combination of $u$ and $d$ quarks contribute to the term ${F}_{nl,n'l'}(q_uB,|q_dB|;r,r')$ with $n=n'=0$, due to the special structure of $\chi^{+}\chi^{+*}$ and $\chi^{-}\chi^{-*}$ in Eq.\eqref{A1}. For charged pion, this kind of combination is absent~\cite{Cao:2019ctl} due to the fact that the LLLs of of $u$ and $d$ quarks cannot form spin singlet at all.
	
	For convenience, we redefine the radii to dimensionless ones $\bar{r}=(eB/2)^{1/2} r$ and $\bar{r}'=(eB/2)^{1/2} r'$, then Eq.\eqref{A2} becomes
	\begin{eqnarray}\label{A3}
	&&	\Pi_{\bar{\rho}^+_1}={-16N_ci\over S}\sum_{n,n'=0}^{n_{\rm max}}\sum_{l=0}^{\mathcal{N}_u}\sum_{l'=0}^{\mathcal{N}_d}\int_{-\infty}^{\infty}\!{dp_0\over2\pi}\int_{-\infty}^{\infty}\!{dp_z\over2\pi}
	{\left({p}_0^{(l-n)+}{p}_0^{(n'-l')-}\!\!-\!p_z^2\!-\!m_um_d\right)\!	{\cal F}_{nl,n'l'}(\tilde{q}_u,|\tilde{q}_d|)\over\left[\left({p}_0^{(l-n)+}\right)^2-(\varepsilon_n^u)^2\right]\left[\left({p}_0^{(n'-l')-}\right)^2-(\varepsilon_{n'}^d)^2\right]},\\
	&&\!\!	{\cal F}_{nl,n'l'}(\tilde{q}_u,|\tilde{q}_d|)={n!n'!\over l!l'!}\tilde{q}_u^{l-n+1}{|\tilde{q}_d|}^{l'-n'+1}\sum_{\bar{r},\bar{r}'}e^{-{\bar{r}^2+(\bar{r}')^2\over 2}}J_{l+l'-n-n'}\left(\bar{r}\bar{r}'\right)(\bar{r}\bar{r}')^{l+l'-n-n'}{F}_{nl,n'l'}(2\tilde{q}_u,2|\tilde{q}_d|;\bar{r},\bar{r}')
	\end{eqnarray}
	with $\tilde{q}=q/e$. It is useful to transform the numerator ${p}_0^{(l-n)+}{p}_0^{(n'-l')-}-p_z^2-m_um_d$ of the integrand in Eq.\eqref{A3} to the following form:
	\bea
	{1\over2}\left[\left({p}_0^{(l-n)+}\right)^2-(\varepsilon_n^u)^2+\left({p}_0^{(n'-l')-}\right)^2-(\varepsilon_{n'}^d)^2-\Omega_{nl,n'l'}^2+(m_u-m_d)^2+2n\,q_uB+2n'|q_dB|\right],
	\eea
	where $\Omega_{nl,n'l'}=(l+l'-n-n'+1)\Omega$. Then the polarization function becomes
	\bea
	\Pi_{\bar{\rho}^+_1}&=&{-8N_ci\over S}\sum_{n,n'=0}^{n_{\rm max}}\sum_{l=0}^{\mathcal{N}_u}\sum_{l'=0}^{\mathcal{N}_d}	{\cal F}_{nl,n'l'}(\tilde{q}_u,|\tilde{q}_d|)\int_{-\infty}^{\infty}\!{dp_0\over2\pi}\int_{-\infty}^{\infty}\!{dp_z\over2\pi}\left\{
	{1\over\left({p}_0^{(l-n)+}\right)^2-(\varepsilon_n^u)^2}+{1\over\left({p}_0^{(n'-l')-}\right)^2-(\varepsilon_{n'}^d)^2}\right.\nonumber\\
	&&\left.+{-\Omega_{nl,n'l'}^2+(m_u-m_d)^2+2n\,q_uB+2n'|q_dB|\over\left[\left({p}_0^{(l-n)+}\right)^2-(\varepsilon_n^u)^2\right]\left[\left({p}_0^{(n'-l')-}\right)^2-(\varepsilon_{n'}^d)^2\right]}\right\}.
	\eea
	Shifting to Euclidean space through the transformations: $p_0\rightarrow i\omega_m$ and $-i\int_{-\infty}^{\infty}\!{dp_0\over2\pi}\rightarrow T\sum_{m=-\infty}^\infty$ and completing the summation over the fermion Matsubara frequency $\omega_m=(2m+1)\pi T$, we have
	\bea
	\Pi_{\bar{\rho}^+_1}&=&{-2N_c\over S}\sum_{n,n'=0}^{n_{\rm max}}\sum_{l=0}^{\mathcal{N}_u}\sum_{l'=0}^{\mathcal{N}_d}	{\cal F}_{nl,n'l'}(\tilde{q}_u,|\tilde{q}_d|)\sum_{s=\pm}\int_{-\infty}^{\infty}{dp_z\over(2\pi)}\left\{\tanh\left({\varepsilon_n^u-s~\Omega_{nl,{1\over2}0}\over2T}\right){1\over\varepsilon_n^u}\Bigg[ 1+\right.\nonumber\\
	&&\left.{-\Omega_{nl,n'l'}^2+(m_u-m_d)^2+2n\,q_uB+2n'|q_dB|\over\left(\varepsilon_n^u-s~\Omega_{nl,n'l'}\right)^2-(\varepsilon_{n'}^d)^2}\Bigg]+\left(\varepsilon_{n'}^d\leftrightarrow\varepsilon_{n}^u,nl\leftrightarrow n'l',q_u\leftrightarrow |q_d|\right)\right\}.
	\eea
	The temperature and rotation dependent part can be separated out as 
	$\Pi_{\bar{\rho}^+_1}-\Pi_{\bar{\rho}^+_1}\big|_{\Omega\rightarrow0, T\rightarrow0},$
	which should be convergent similar to that of charged pion~\cite{Cao:2019ctl}. Here, the subtracting term is just the polarization function in pure magnetic field, which has been regularized in Ref.~\cite{Cao:2019res} as $\left[\Pi_{\bar{\rho}^+_1}^B-\Pi_{\bar{\rho}^+_1}^{o(B^2)}\right]+\Pi_{\bar{\rho}^+_1}^\Lambda$. We close this section by listing the relevant terms:
		\begin{eqnarray}
	\Pi_{\bar{\rho}^+_1}^B&=&-{N_c\over8\pi^2}\!\!\int\!{\di s\over s}\!\!\int_{-1}^1\!\! {\di u}~{e^{-s\left[m_{\rm u}^2u^+\!+m_{\rm d}^2u^-\right]}}\!\!\left(m_{\rm u}m_{\rm d}\!+\!{1\over s}\right)\!\!{\left[1\!+\!\tanh\left(q_{\rm u}Bsu^+\right)\right]\left[1\!-\!\tanh\left(q_{\rm d}Bsu^-\right)\right]\over {\tanh\left(q_{\rm u}Bsu^+\right)\over q_{\rm u}Bs}+{\tanh\left(q_{\rm d}Bsu^-\right)\over q_{\rm d}Bs}}
	\end{eqnarray}
	with $u^\pm=(1\pm u)/2$, $\Pi_{\bar{\rho}^+_1}^{o(B^2)}$ is the weak $B$ expansion of $\Pi_{\bar{\rho}^+_1}^B$ to order $o(B^2)$ and the term with three-momentum cutoff $\Lambda$ is
\begin{eqnarray}
\Pi_{\bar{\rho}^+_1}^\Lambda&=&-N_c\int_0^\Lambda\!\! {k^2dk\over\pi^2}\frac{(E_{\rm u}E_{\rm d}\!+\!{m_{\rm u}}{m_{\rm d}}\!+\!{1\over3}k^2)}{E_{\rm u}E_{\rm d}(E_{\rm u}\!+\!E_{\rm d})}-{N_c}\int_0^\Lambda {k^2dk\over2\pi^2}\left\{{{q_{\rm u}B}\over(E_{\rm u}\!+\!E_{\rm d})^2}\left[\left(\frac{E_{\rm u}E_{\rm d}\!+\!{m_{\rm u}}{m_{\rm d}}\!+\!{1\over3}k^2}{E_{\rm u}^3}\!+\!{1\over E_{\rm u}}\!+\!{1\over E_{\rm d}}\right)\right.\right.\nonumber\\
&&\left.\left.-\frac{(m_{\rm u}\!-\!m_{\rm d})^2\!+\!{4\over3}k^2}{E_{\rm u}^2E_{\rm d}}\right]\!-\!(u\leftrightarrow d)\right\}.
\end{eqnarray}

\subsection{For Schwinger phase in curved space}\label{ASPC}
Next, we choose the Schwinger phase of the form in curved space, that is, $\Phi_M={eB\over2}\sin[\theta-\theta'+\Omega(t-t')]rr'$. By taking variable transformation of the angle: $\theta-\theta'\rightarrow \theta-\theta'-\Omega(t-t')$, we find that the corresponding polarization function can be modified from Eq.\eqref{A3} by changing ${p}_0^{(l-n)+}$ to ${p}_0^{(n'-l')+}$, that is,
\begin{eqnarray}\label{A4}
&&	\Pi_{\bar{\rho}^+_1}={-16N_ci\over S}\sum_{n,n'=0}^{n_{\rm max}}\sum_{l=0}^{\mathcal{N}_u}\sum_{l'=0}^{\mathcal{N}_d}\int_{-\infty}^{\infty}\!{dp_0\over2\pi}\int_{-\infty}^{\infty}\!{dp_z\over2\pi}
{\left({p}_0^{(n'-l')+}{p}_0^{(n'-l')-}\!\!-\!p_z^2\!-\!m_um_d\right)	{\cal F}_{nl,n'l'}(\tilde{q}_u,|\tilde{q}_d|)\over\left[\left({p}_0^{(n'-l')+}\right)^2-(\varepsilon_n^u)^2\right]\left[\left({p}_0^{(n'-l')-}\right)^2-(\varepsilon_{n'}^d)^2\right]}.
\end{eqnarray}
Then, in a similar process as the previous section, the summation over the Fermion Matsubara frequency gives
\bea
\Pi_{\bar{\rho}^+_1}&=&{-2N_c\over S}\sum_{n,n'=0}^{n_{\rm max}}\sum_{l=0}^{\mathcal{N}_u}\sum_{l'=0}^{\mathcal{N}_d}	{\cal F}_{nl,n'l'}(\tilde{q}_u,|\tilde{q}_d|)\sum_{s=\pm}\int_{-\infty}^{\infty}{dp_z\over(2\pi)}\Bigg\{\tanh\left({\varepsilon_n^u\!+\!s~\Omega_{{3\over2}0,n'l'}\over2T}\right){1\over\varepsilon_n^u}+\tanh\left({\varepsilon_{n'}^d-s~\Omega_{{1\over2}0,n'l'}\over2T}\right){1\over\varepsilon_{n'}^d}\nonumber\\
&&+\left[{-\Omega^2\!+\!(m_u\!-\!m_d)^2\!+\!2n\,q_uB\!+\!2n'|q_dB|}\right]\left[{1\over\varepsilon_n^u}{\tanh\left({\varepsilon_n^u\!+\!s~\Omega_{{3\over2}0,n'l'}\over2T}\right)\over\left(\varepsilon_n^u-s\,\Omega\right)^2-(\varepsilon_{n'}^d)^2}+{1\over\varepsilon_{n'}^d}{\tanh\left({\varepsilon_{n'}^d-s~\Omega_{{1\over2}0,n'l'}\over2T}\right)\over\left(\varepsilon_{n'}^d-s\,\Omega\right)^2-(\varepsilon_n^u)^2}\right]\Bigg\}.
\eea
In contrary to that of charged pion~\cite{Chen:2019tcp}, the angular velocity $\Omega$ can still plays a role of effective isospin chemical potential to $\bar{\rho}^+_1$ in this case, see the denominators. Finally, one should keep in mind that the regularization to Eq.\eqref{A4} is performed in the same way as that of SPM. 
\end{widetext}

\section{Possibility of charged rho superconductor}\label{numerical}
In order to carry out numerical calculations, we choose the following parameters for the scalar-pseudoscalar sector: $m_{\rm u}=m_{\rm d}=5.5~{\rm MeV}, m_{\rm s}=140.7~{\rm MeV}, \Lambda=602.3~{\rm MeV}, G_S\Lambda^2=1.835$ and $K\Lambda^5=12.36$~\cite{Rehberg:1995kh}. To avoid artifacts, the vector coupling constant is fixed to $G_V\Lambda^2=2.527$ by fitting to the vacuum mass of $\rho$ meson: $m_\rho^v=0.7~{\rm GeV}$~\cite{Cao:2019res}. Following the study of Ref.~\cite{Cao:2019res}, we consider a cylindrical system with the radius $R=20/\sqrt{eB}$ and constrain the rotation by $\Omega R\leq1$ for causality.

\begin{figure}[!htb]
	\centering
	\includegraphics[width=0.42\textwidth]{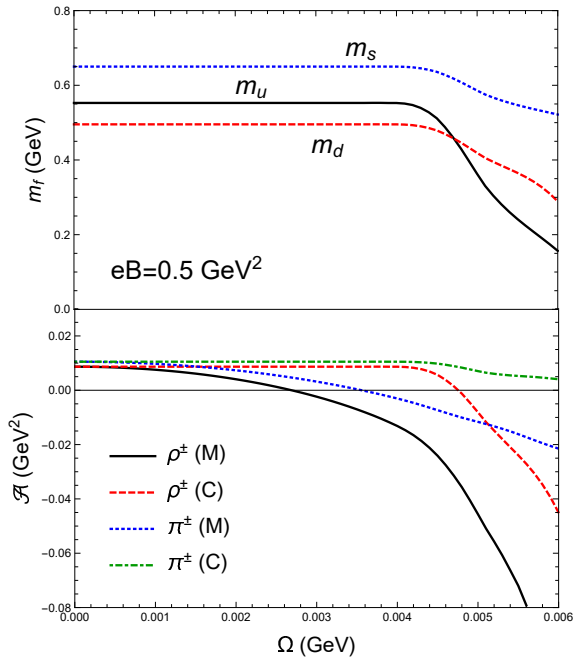}
	\caption{The evolutions of the dynamical quark masses $m_{\rm f}$ (upper panel) and the quadratic GL expansion coefficients ${\cal A}$ (lower panel) with the angular velocity $\Omega$ at the magnetic field $eB=0.5~{\rm GeV}^2$. In the lower panel, the coefficients of charged rho condensation are compared to those of charged pion for both choices of SPM and SPC, denoted by "M" and "C", respectively.}\label{B05}
\end{figure}

\begin{figure}[!htb]
	\centering
	\includegraphics[width=0.42\textwidth]{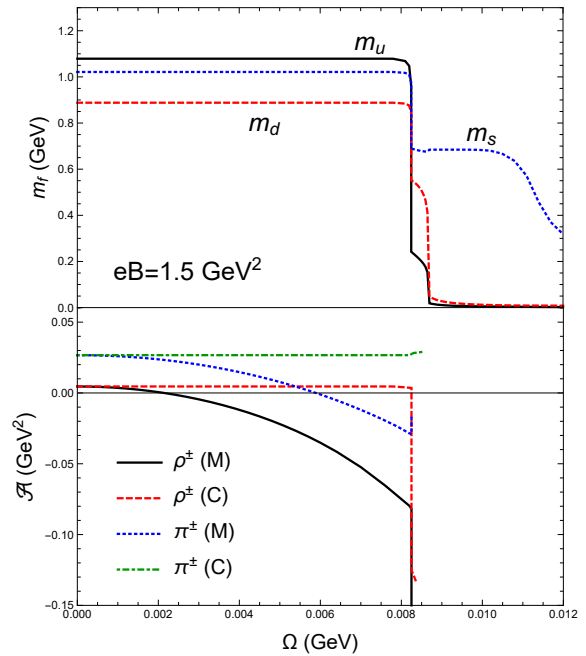}
	\caption{The evolutions of the dynamical quark masses $m_{\rm f}$ (upper panel) and the quadratic GL expansion coefficients ${\cal A}$ (lower panel) with the angular velocity $\Omega$ at the magnetic field $eB=1.5~{\rm GeV}^2$. The conventions are the same as in Fig.\ref{B05}.}\label{B15}
\end{figure}

We choose two strong enough magnetic fields for illumination: $eB=0.5~{\rm GeV}^2$ and $eB=1.5~{\rm GeV}^2$, which are on different sides of the minimum point of the $\bar{\rho}^+_1$ mass found in our previous work~\cite{Cao:2019res}. The numerical results are shown in Figs.\ref{B05} and \ref{B15}, respectively. As can be seen in the upper panels, the quark masses all decrease with $\Omega$ in both cases, but the chiral symmetry restoration ($\chi$SR) shows a crossover feature for $eB=0.5~{\rm GeV}^2$ and a first-order one for $eB=1.5~{\rm GeV}^2$. Along with the $\chi$SR, the quadratic GL expansion coefficients are evaluated for charged rho meson with both choices of SPM and SPC, see the lower panels in comparison with those of charged pions. For either choice of Schwinger phase, the CRS can always happen and is favored over the CPS with large enough $\Omega$. 

We've checked for $eB=0.5~{\rm GeV}^2$ that the CPS is indeed disfavored with SPM at large $\Omega$ ($\geq 0.026~{\rm GeV}$) thus qualitatively consistent with that found in Ref.~\cite{Chen:2019tcp}. But for charged rho meson, the quadratic coefficient keeps decreasing to
an order of $-100~{\rm GeV}^2$ at $\Omega=0.026~{\rm GeV}$ without any signature of turning up. The discontinuity in Fig.\ref{B15} seems to contradict with the continuous transition assumption in the GL approach, but it surely demonstrates an instability to the $\chi$SR phase. The situation might be similar to that of diquark condensation at the critical baryon chemical potential, so here can probably be a first-order transition to CRS. In the lower panel of Fig.\ref{B15}, one note that the coefficients ${\cal A}$ increase for charged pion but decrease for charged rho at the critical $\Omega\ (\sim8.25~{\rm MeV})$. At last, though not illuminated in the plots, it has been checked that the charged axial vector condensation might be favored over $\chi$SR or CPS phase at large enough $\Omega$ but never over CRS phase.

\section{Conclusions and discussions}\label{conclusions}
In this work, the possibility of charged rho superconductor in the presence of parallel magnetic field and rotation was intuitively studied in Weinberg model and extensively explored within $SU(3)$ Nambu--Jona-Lasino model. The charged $\pi$ and $\rho$ condensations were both well convinced in the point particle picture. By following similar schemes as the previous works on CPS~\cite{Cao:2019ctl,Chen:2019tcp}, the CRS was found to be favored over chiral symmetry breaking,  $\chi$SR  and CPS phases at large $\Omega$, for both choices of Schwinger phase in Minkovski and curved spaces. As the chiral partner of $\rho$ mesons, the charged axial vector meson was even checked in advance; and it turned out that CRS is still robust against that at large $\Omega$. Indeed, the NJL model study qualitatively supports the intuitions about the rotation effect on mesons in the point particle approximation. 

In the future, more realistic but complicated study will be performed to looking for the true ground state of QCD system in PMR by taking into account the boundary condition and inhomogeneous forms of condensates consistently~\cite{Chen:2017xrj,Wang:2019nhd}. As discussed in Sec.\ref{Weinberg model}, the effective regime of $l$ should be determined by the total energy self-consistently and can be much greater than $\mathcal{N}$ for large $\Omega$. In this case, the CPS or CRS is expected to emerge at a smaller threshold of $\Omega$ compared to what we found here in NJL model. One should notice that: Though CRS is more favored for the chosen magnetic fields, there is still a window of $\Omega$ for CPC phase when $B$ is relatively weak. Eventually, as the PMR is relevant to the circumstance in peripheral heavy ion collisions, it will be interesting to explore the possible signatures for the competitions among $\chi$SR, CPS and CRS in experiments.

\emph{Acknowledgments}---
We thank Jinfeng Liao for helpful discussions. G.C. is supported by the National Natural Science Foundation of China with Grant No. 11805290 and Young Teachers Training Program of Sun Yat-sen University with Grant No. 19lgpy282.

\end{document}